\documentclass[aip,apl,reprint,groupedaddress]{revtex4-1}
\usepackage{graphics}
\usepackage{natbib}
\usepackage{amsmath}
\newcommand{\Tone}{\ensuremath{{T}_1}}
\begin{document}

\title{Bulk and surface loss in superconducting transmon qubits}

\author{Oliver Dial, Douglas T. McClure, Stefano Poletto\dag, Jay M. Gambetta, David W. Abraham, Jerry M. Chow, Matthias Steffen}
\affiliation{IBM T.J. Watson Research Center, Yorktown Heights, NY  10598, USA\\
\dag: Now QuTech Advanced Research Center and
Kavli Institute of Nanoscience, Delft University of Technology,
Lorentzweg 1, 2628 CJ Delft, The Netherlands}
\date{\today}

\begin{abstract}Decoherence of superconducting transmon qubits is purported to be consistent with surface loss from two-level systems on the substrate surface. Here, we present a study of surface loss in transmon devices, explicitly designed to have varying sensitivities to different surface loss contributors. Our experiments also encompass two particular different sapphire substrates, which reveal the onset of a yet unknown additional loss mechanism outside of surface loss for one of the substrates. Tests across different wafers and devices demonstrate substantial variation, and we emphasize the importance of testing large numbers of devices for disentangling different sources of decoherence. 
\end{abstract}
\pacs{}

\maketitle

Superconducting transmon qubits~\cite{koch07}, often studied in both planar (2D) and waveguide (3D) modalities, have become important elements for extensible quantum computing architectures~\cite{Kelly:2015aa,Corcoles:2015aa,Riste:2015aa} which employ circuit quantum electrodynamics (cQED). Typically, transmon coherence times are largely dominated by amplitude damping ($T_1$), and for low temperatures and devices which are not Purcell limited~\cite{Houck2008}, are often observed to become shorter with smaller physical device sizes~\cite{Chang2013,Barends2013}. This trend, complementary to that observed for quality factor dependence on physical trace and spacing widths in planar resonators~\cite{wang09a,Geerlings2012}, is suggestive that devices are surface loss limited; as the transmon becomes smaller, a greater fraction of the electric field energy lies within a hypothetical layer of lossy material on different device surfaces\cite{gao08,wenner11,quintana14}.

The 3D waveguide approach to cQED~\cite{paik11} is particularly well-suited for investigation of loss mechanisms directly on fabricated transmon devices, because of the clean microwave environment it provides. A transmon qubit device is only subject to environmental coupling through the high-quality rectangular waveguide modes supported by the 3D cavity. As such, the Purcell limit can be high, providing a window into the losses due to the qubit design, fabrication processes, and bulk materials. These particular losses, we refer to as \emph{intrinsic} to the devices.

In this Letter, we present an experimental investigation of intrinsic loss
mechanisms of transmon devices exploiting 3D waveguide approach. Surface losses
of different transmon device interfaces are examined via targeted designs with
varying participation susceptibilities to these losses. Our results are
consistent with the substrate-vacuum and substrate-metal interfaces being the
most likely contributors to transmon decoherence. Fabrication processing
variations in the device lift-off medium to target removal of lossy interfaces
are also studied, though with no strong conclusion. Finally, we observe a strong
overall dependence on the intrinsic bulk loss properties of the substrate which
the qubits are fabricated on, indicating the superiority of heat exchanger
method (HEM) sapphire, over the more commonly used edge-defined film-fed grown (EFG) sapphire.

\begin{figure}
\resizebox{\columnwidth}{!}{\includegraphics{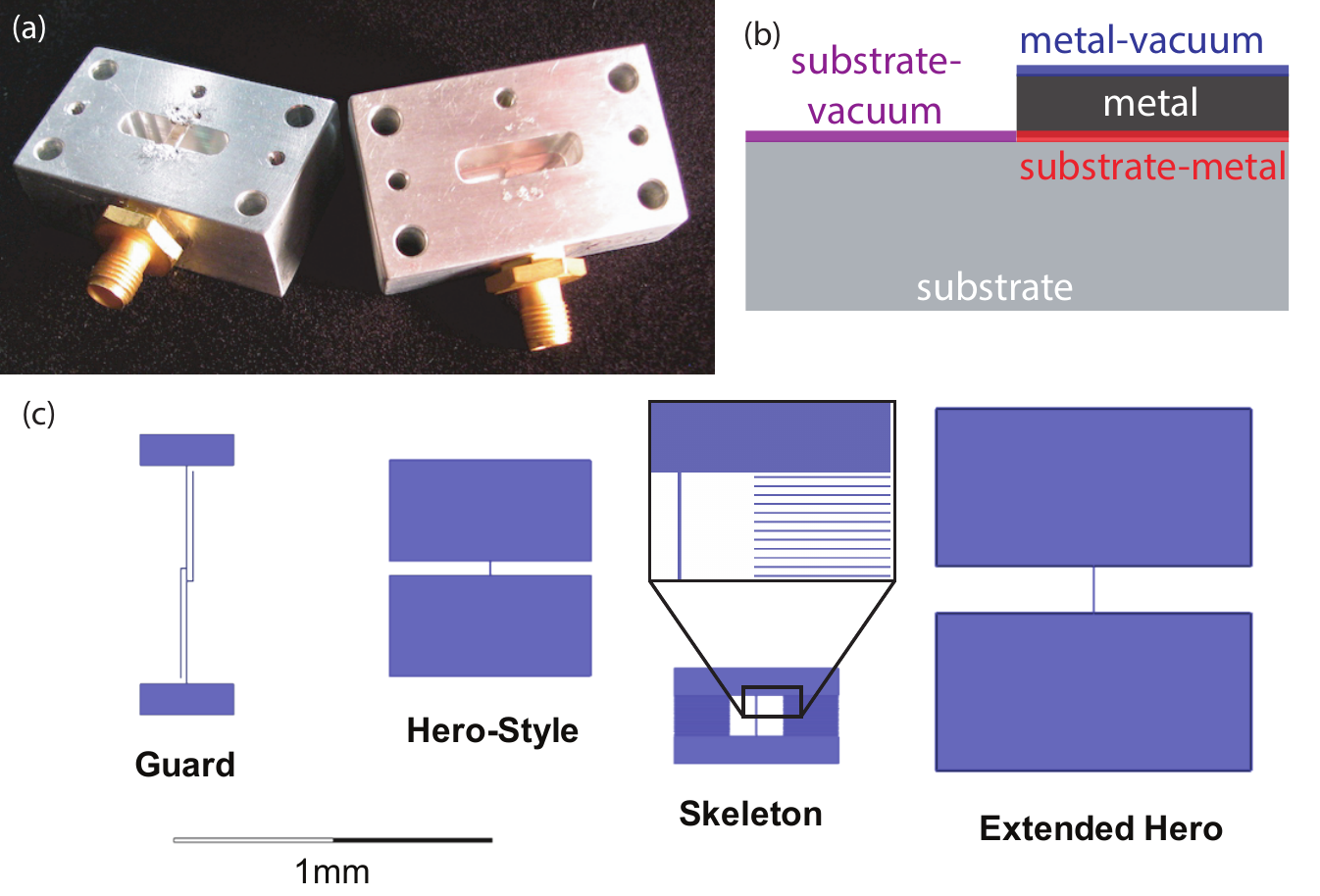}}
\caption{(color online) (a) A typical aluminum cavity as used in this work, housing a single 3D transmon device for test. (b) Layered cartoon of transmon device layers, along with targeted lossy interface layers. For our loss simulations, participation ratios in the substrate-metal, substrate-vacuum, and metal-vacuum layers are estimated. (c) The four reference transmon qubit designs studied, with varying susceptibility to the lossy interface layers.
\label{transmons}
}
\end{figure}

First, we formalize the surface loss picture in the transmon devices by examining the participation of electric fields in various lossy interface layers. We consider a thin surface layer contained in a volume $V_s$, from which we can define the electric participation ratio  $R=\iiint \vec{E} \cdot \vec{D} dV_s / E_{\text{tot}}$, where $E_{\text{tot}}$ is the total electrical energy $\iiint \vec{E} \cdot \vec{D} dV$. If this layer, with a small microwave loss tangent $\delta$, is the most dominant source of loss, the qubit will then have a quality factor  $Q \equiv \omega \Tone = 1/(R \tan\delta)$ where $\omega$ is the qubit frequency.  Finite element solvers (Ansoft HFSS) can be used to estimate the surface participation of various films in complex geometries (such as for transmon qubits). However, there is a vast difference in scale between typically estimated thicknesses of the thin surface loss layers ($\sim$nm) and the width-scale typical of 3D qubit geometries ($>$100 $\mu$m) which renders this computation difficult.  To counteract this, we make the assumption that the electric field variation is small across the thickness, $t$, of a lossy layer, such that we can approximate the volume integral with a surface integral, giving
$R = \iint t \vec{E}_{c} \cdot \vec{D}_c dS/E_{tot}$. Here the surface of
integration is the hypothetical lossy layer and  the integral is over the volume
of the device.   In the absence of any outside knowledge of the nature of the
lossy layer, we note that $Q^{-1} \propto t \delta$, so $T_1$  measurements
alone cannot determine $t$ or $\delta$, but only their product.  We thus
introduce the surface loss sensitivity $r \equiv R/t$ which has units of inverse
length and describes how sensitive a device is to a given surface.

To investigate surface loss in 3D superconducting transmon qubit
devices\cite{paik11} [Fig.~\ref{transmons}(a)] we hypothesize that each of the
qubit's important interfaces, substrate-vacuum, substrate-metal, and
metal-vacuum, may have a uniform lossy layer covering it [see
Fig.~\ref{transmons}(b)].  We designed four benchmark transmon qubits designed to have different sensitivities to these interfaces which are shown in Fig.~\ref{transmons}(c). The ``Hero" and ``Extended Hero" are motivated by the original 3D qubit work of Ref.~\citenum{paik11}, but reflect a scaling to minimize overall sensitivity to all surface losses in the larger design. The ``Guard" design maintains a small gap throughout the extent of the qubit, so as to be intentionally very sensitive to both the substrate-metal and substrate-vacuum layers. Finally, the ``Skeleton" design actually contains many thin floating islands of metal in the gap between the two main qubit capacitive pads. These additional ``bones'' in the Skeleton design are meant to not contribute to the overall capacitance, but in effect to minimize the sensitivity of the qubit to the substrate-vacuum but more sensitive to the substrate-metal interface.

\begin{figure}
\resizebox{\columnwidth}{!}{\includegraphics{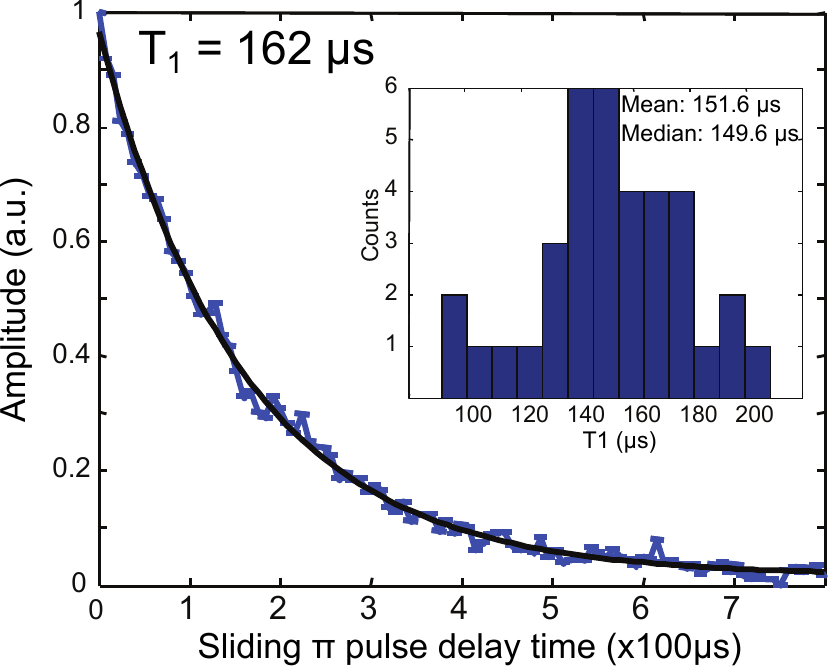}}
\caption{Typical $T_1$ trace from an HEM sapphire Extended Hero qubit; histogram distribution of $T_1$ measured across several hours. \label{t1}}
\end{figure}

We fabricated and measured multiple instances of these different qubit designs
both within the same wafers and between wafers on two types of substrates:
c-Plane EFG sapphire from Kyocera, used in previous work\cite{rigetti12}, and
HEM Sapphire from GT Advanced Technologies. We note that all of our results are
dependent on accurate and reproducible measurements of qubit $T_1$, a quantity observed to fluctuate over time.  Each $T_1$ trace is averaged across several hours to average across this fluctuation, and the mean value is reported in this work.  A typical trace and histogram across several hours is shown in Fig.~\ref{t1}(a-b) for a particular Extended Hero qubit on an HEM sapphire substrate.

To visualize the importance that loss at different interfaces may have, we plot
the quality factor $Q$ of the measured qubits against $1/r$ for each particular
surface layer; a linear relationship would be expected if the $T_1$ of the qubit
were limited by that layer.  The dielectric constant of the imputed layer also
has some impact on the resulting loss sensitivity.  We assume a value of 6.2, 
intermediate between that typical for oxides and for organic films, although the precise value does not materially impact our conclusions.  As part of this study, note that we have prepared a number of qubits with identical capacitor designs but different Josephson junction areas, resulting in qubit frequencies that differed by a factor of two. Across these different devices, we find $Q$ to be similar, with $T_1$ being overall shorter for the higher frequency devices, consistent with the loss model indicated above. This confirms that the correct way to compare qubits of different frequency in our experiment is by comparing $Q$, not $T_1$.  Our measurements are split across two dilution refrigerators, with microwave switches and multiple input lines allowing measurements of up to twelve devices in a single cooldown. Using long $T_1$ pre-vetted ``canary'' qubits shared between the refrigerators and for different input lines, we have confirmed that there is no significant variation for any of the test setups.

\begin{figure*}
\resizebox{\textwidth}{!}{\includegraphics{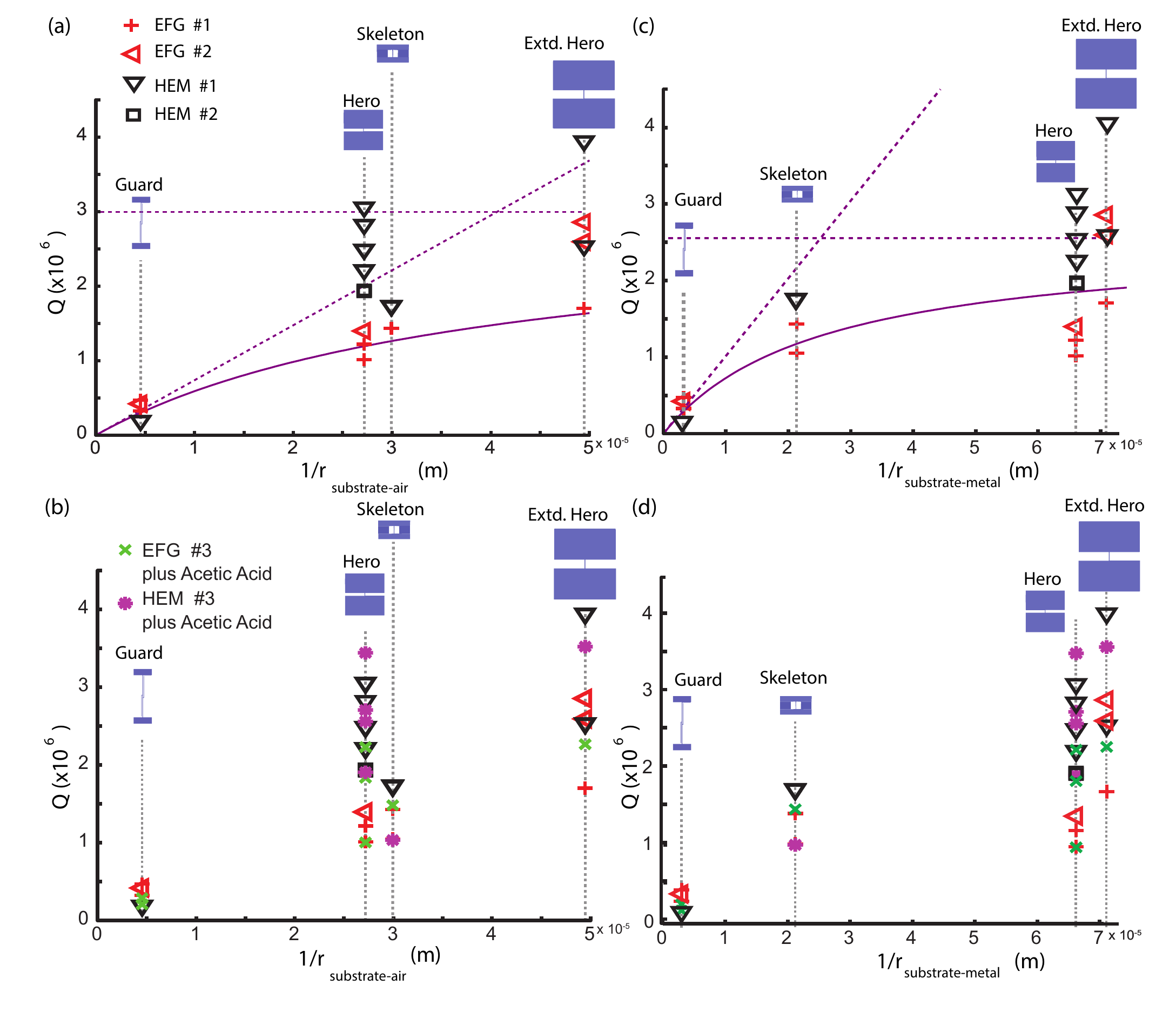}}
\caption{(a) $Q$ versus inverse substrate-vacuum loss sensitivity for several qubits on two EFG sapphire (red) and HEM sapphire (black) wafers.  A simple surface loss model appears as a straight line through the origin, while a horizontal line is indicative of a fixed background loss (dashed purple lines).  Each point represents a single qubit, and measurement error is small compared to qubit-to-qubit variation within the same process condition.  A model with both the combined surface loss and a background loss best fit to the EFG data is shown as a solid purple line. (b) Results for qubits processed with glacial acetic acid are indistinguishable from those processed with acetone.  (c,d) Similar conclusions are drawn if devices are compared to the substrate-metal participation.  These data cannot resolve if substrate-metal or substrate-vacuum loss dominates, or if each is equally important. \label{lossplots}}
\end{figure*}

The extracted $Q$ values of 35 qubit devices fabricated on 6 wafers are shown in
Fig.~\ref{lossplots}.  Each position along the $x$-axis corresponds to one of
the four qubit types, each distinct symbol represents devices drawn from a
different wafer, and each color indicates a different fabrication process
variation. In Fig.~\ref{lossplots}(a), red points indicate qubits fabricated on
EFG sapphire substrates and black points indicate HEM sapphire.  For the EFG
data, we note that although the dependence of $Q$ on the substrate-vacuum
participation ratio is monotonic, it saturates at very large qubit designs
(small surface sensitivities) to a fixed value. The solid purple curve is a
hypothesized fit to this with two contributions: a surface loss proportional to
$1/r$ and a fixed background loss (dashed purple lines). This trend is
 best-fit to $r$ of
$1.6\times10^{-11}/\mathrm{m}$ and the fixed background $Q$ of 3 million.  By comparison, if we examine the substrate-metal interface as in
Fig.~\ref{lossplots}(c), we also find a saturation at small participation.
Within the scatter of our data we are not able to rule out either
substrate-metal or substrate-vacuum as dominating the loss.

Taken together, the observed fixed background loss and wafer-to-wafer
variability were suggestive of substrate loss, i.e. that the sapphire substrates
had a fixed, non-neglible loss tangent of around $3\times10^{-7}$. Particular
pieces of specially prepared HEM sapphire have been known to have very low
cryogenic loss tangents\cite{creedon11}.  To elucidate this issue, we prepared
additional 3D qubits on HEM sapphire, and observed improved quality factors for
the Hero and Extended Hero design (black data in Fig.~\ref{lossplots} a, c).
Based on this result, we believe the original Hero and Extended Hero devices on
EFG sapphire were largely limited by bulk substrate loss, which would be
independent of qubit design.

As a final process variation, we used glacial acetic acid for liftoff rather
than acetone. This is reported to give cleaner surfaces when processing
graphene, another surface sensitive material\cite{chavez03,her13}.  These
additional data are shown in green and purple for EFG and HEM substrates
respectively in Fig.~\ref{lossplots}(b,d); there appears to be no large change as compared to our original processing.  We note that given our observed spreads, large numbers of devices need to be compared both within a wafer and between wafers to reliably resolve even large effects.  Measuring single qubits is inadequate.

The results in this work are likely to be particular to the details of process
steps and conditions currently in use at our fabrication facility; however, the
approach of designing a series of qubits of carefully controlled loss
sensitivities as a technique for disentangling multiple coexisting loss
mechanisms as well as extrapolating practical limits on qubit coherence times
in various material systems is broadly applicable in both 2D and 3D qubits.

The authors would like to acknowledge discussions and contributions from George 	A. Keefe, Mary B. Rothwell. We acknowledge support from IARPA under contract W911NF-10-1-0324.

\bibliography{main}

\begin{thebibliography}{17}%
\makeatletter
\providecommand \@ifxundefined [1]{%
 \@ifx{#1\undefined}
}%
\providecommand \@ifnum [1]{%
 \ifnum #1\expandafter \@firstoftwo
 \else \expandafter \@secondoftwo
 \fi
}%
\providecommand \@ifx [1]{%
 \ifx #1\expandafter \@firstoftwo
 \else \expandafter \@secondoftwo
 \fi
}%
\providecommand \natexlab [1]{#1}%
\providecommand \enquote  [1]{``#1''}%
\providecommand \bibnamefont  [1]{#1}%
\providecommand \bibfnamefont [1]{#1}%
\providecommand \citenamefont [1]{#1}%
\providecommand \href@noop [0]{\@secondoftwo}%
\providecommand \href [0]{\begingroup \@sanitize@url \@href}%
\providecommand \@href[1]{\@@startlink{#1}\@@href}%
\providecommand \@@href[1]{\endgroup#1\@@endlink}%
\providecommand \@sanitize@url [0]{\catcode `\\12\catcode `\$12\catcode
  `\&12\catcode `\#12\catcode `\^12\catcode `\_12\catcode `\%12\relax}%
\providecommand \@@startlink[1]{}%
\providecommand \@@endlink[0]{}%
\providecommand \url  [0]{\begingroup\@sanitize@url \@url }%
\providecommand \@url [1]{\endgroup\@href {#1}{\urlprefix }}%
\providecommand \urlprefix  [0]{URL }%
\providecommand \Eprint [0]{\href }%
\providecommand \doibase [0]{http://dx.doi.org/}%
\providecommand \selectlanguage [0]{\@gobble}%
\providecommand \bibinfo  [0]{\@secondoftwo}%
\providecommand \bibfield  [0]{\@secondoftwo}%
\providecommand \translation [1]{[#1]}%
\providecommand \BibitemOpen [0]{}%
\providecommand \bibitemStop [0]{}%
\providecommand \bibitemNoStop [0]{.\EOS\space}%
\providecommand \EOS [0]{\spacefactor3000\relax}%
\providecommand \BibitemShut  [1]{\csname bibitem#1\endcsname}%
\let\auto@bib@innerbib\@empty
\bibitem [{\citenamefont {Koch}\ \emph {et~al.}(2007)\citenamefont {Koch},
  \citenamefont {Yu}, \citenamefont {Gambetta}, \citenamefont {Houck},
  \citenamefont {Schuster}, \citenamefont {Majer}, \citenamefont {Blais},
  \citenamefont {Devoret}, \citenamefont {Girvin},\ and\ \citenamefont
  {Schoelkopf}}]{koch07}%
  \BibitemOpen
  \bibfield  {author} {\bibinfo {author} {\bibfnamefont {J.}~\bibnamefont
  {Koch}}, \bibinfo {author} {\bibfnamefont {T.~M.}\ \bibnamefont {Yu}},
  \bibinfo {author} {\bibfnamefont {J.}~\bibnamefont {Gambetta}}, \bibinfo
  {author} {\bibfnamefont {A.~A.}\ \bibnamefont {Houck}}, \bibinfo {author}
  {\bibfnamefont {D.~I.}\ \bibnamefont {Schuster}}, \bibinfo {author}
  {\bibfnamefont {J.}~\bibnamefont {Majer}}, \bibinfo {author} {\bibfnamefont
  {A.}~\bibnamefont {Blais}}, \bibinfo {author} {\bibfnamefont {M.~H.}\
  \bibnamefont {Devoret}}, \bibinfo {author} {\bibfnamefont {S.~M.}\
  \bibnamefont {Girvin}}, \ and\ \bibinfo {author} {\bibfnamefont {R.~J.}\
  \bibnamefont {Schoelkopf}},\ }\href {\doibase 10.1103/PhysRevA.76.042319}
  {\bibfield  {journal} {\bibinfo  {journal} {Physical Review A}\ }\textbf
  {\bibinfo {volume} {76}},\ \bibinfo {pages} {042319} (\bibinfo {year}
  {2007})}\BibitemShut {NoStop}%
\bibitem [{\citenamefont {Kelly}\ \emph {et~al.}(2015)\citenamefont {Kelly},
  \citenamefont {Barends}, \citenamefont {Fowler}, \citenamefont {Megrant},
  \citenamefont {Jeffrey}, \citenamefont {White}, \citenamefont {Sank},
  \citenamefont {Mutus}, \citenamefont {Campbell}, \citenamefont {Chen},
  \citenamefont {Chen}, \citenamefont {Chiaro}, \citenamefont {Dunsworth},
  \citenamefont {Hoi}, \citenamefont {Neill}, \citenamefont {O'Malley},
  \citenamefont {Quintana}, \citenamefont {Roushan}, \citenamefont
  {Vainsencher}, \citenamefont {Wenner}, \citenamefont {Cleland},\ and\
  \citenamefont {Martinis}}]{Kelly:2015aa}%
  \BibitemOpen
  \bibfield  {author} {\bibinfo {author} {\bibfnamefont {J.}~\bibnamefont
  {Kelly}}, \bibinfo {author} {\bibfnamefont {R.}~\bibnamefont {Barends}},
  \bibinfo {author} {\bibfnamefont {A.~G.}\ \bibnamefont {Fowler}}, \bibinfo
  {author} {\bibfnamefont {A.}~\bibnamefont {Megrant}}, \bibinfo {author}
  {\bibfnamefont {E.}~\bibnamefont {Jeffrey}}, \bibinfo {author} {\bibfnamefont
  {T.~C.}\ \bibnamefont {White}}, \bibinfo {author} {\bibfnamefont
  {D.}~\bibnamefont {Sank}}, \bibinfo {author} {\bibfnamefont {J.~Y.}\
  \bibnamefont {Mutus}}, \bibinfo {author} {\bibfnamefont {B.}~\bibnamefont
  {Campbell}}, \bibinfo {author} {\bibfnamefont {Y.}~\bibnamefont {Chen}},
  \bibinfo {author} {\bibfnamefont {Z.}~\bibnamefont {Chen}}, \bibinfo {author}
  {\bibfnamefont {B.}~\bibnamefont {Chiaro}}, \bibinfo {author} {\bibfnamefont
  {A.}~\bibnamefont {Dunsworth}}, \bibinfo {author} {\bibfnamefont {I.~C.}\
  \bibnamefont {Hoi}}, \bibinfo {author} {\bibfnamefont {C.}~\bibnamefont
  {Neill}}, \bibinfo {author} {\bibfnamefont {P.~J.~J.}\ \bibnamefont
  {O'Malley}}, \bibinfo {author} {\bibfnamefont {C.}~\bibnamefont {Quintana}},
  \bibinfo {author} {\bibfnamefont {P.}~\bibnamefont {Roushan}}, \bibinfo
  {author} {\bibfnamefont {A.}~\bibnamefont {Vainsencher}}, \bibinfo {author}
  {\bibfnamefont {J.}~\bibnamefont {Wenner}}, \bibinfo {author} {\bibfnamefont
  {A.~N.}\ \bibnamefont {Cleland}}, \ and\ \bibinfo {author} {\bibfnamefont
  {J.~M.}\ \bibnamefont {Martinis}},\ }\href
  {http://dx.doi.org/10.1038/nature14270} {\bibfield  {journal} {\bibinfo
  {journal} {Nature}\ }\textbf {\bibinfo {volume} {519}},\ \bibinfo {pages}
  {66} (\bibinfo {year} {2015})}\BibitemShut {NoStop}%
\bibitem [{\citenamefont {Corcoles}\ \emph {et~al.}(2015)\citenamefont
  {Corcoles}, \citenamefont {Magesan}, \citenamefont {Srinivasan},
  \citenamefont {Cross}, \citenamefont {Steffen}, \citenamefont {Gambetta},\
  and\ \citenamefont {Chow}}]{Corcoles:2015aa}%
  \BibitemOpen
  \bibfield  {author} {\bibinfo {author} {\bibfnamefont {A.~D.}\ \bibnamefont
  {Corcoles}}, \bibinfo {author} {\bibfnamefont {E.}~\bibnamefont {Magesan}},
  \bibinfo {author} {\bibfnamefont {S.~J.}\ \bibnamefont {Srinivasan}},
  \bibinfo {author} {\bibfnamefont {A.~W.}\ \bibnamefont {Cross}}, \bibinfo
  {author} {\bibfnamefont {M.}~\bibnamefont {Steffen}}, \bibinfo {author}
  {\bibfnamefont {J.~M.}\ \bibnamefont {Gambetta}}, \ and\ \bibinfo {author}
  {\bibfnamefont {J.~M.}\ \bibnamefont {Chow}},\ }\href
  {http://dx.doi.org/10.1038/ncomms7979} {\bibfield  {journal} {\bibinfo
  {journal} {Nat Commun}\ }\textbf {\bibinfo {volume} {6}} (\bibinfo {year}
  {2015})}\BibitemShut {NoStop}%
\bibitem [{\citenamefont {Riste}\ \emph {et~al.}(2015)\citenamefont {Riste},
  \citenamefont {Poletto}, \citenamefont {Huang}, \citenamefont {Bruno},
  \citenamefont {Vesterinen}, \citenamefont {Saira},\ and\ \citenamefont
  {DiCarlo}}]{Riste:2015aa}%
  \BibitemOpen
  \bibfield  {author} {\bibinfo {author} {\bibfnamefont {D.}~\bibnamefont
  {Riste}}, \bibinfo {author} {\bibfnamefont {S.}~\bibnamefont {Poletto}},
  \bibinfo {author} {\bibfnamefont {M.~Z.}\ \bibnamefont {Huang}}, \bibinfo
  {author} {\bibfnamefont {A.}~\bibnamefont {Bruno}}, \bibinfo {author}
  {\bibfnamefont {V.}~\bibnamefont {Vesterinen}}, \bibinfo {author}
  {\bibfnamefont {O.~P.}\ \bibnamefont {Saira}}, \ and\ \bibinfo {author}
  {\bibfnamefont {L.}~\bibnamefont {DiCarlo}},\ }\href
  {http://dx.doi.org/10.1038/ncomms7983} {\bibfield  {journal} {\bibinfo
  {journal} {Nat Commun}\ }\textbf {\bibinfo {volume} {6}} (\bibinfo {year}
  {2015})}\BibitemShut {NoStop}%
\bibitem [{\citenamefont {Houck}\ \emph {et~al.}(2008)\citenamefont {Houck},
  \citenamefont {Schreier}, \citenamefont {Johnson}, \citenamefont {Chow},
  \citenamefont {Koch}, \citenamefont {Gambetta}, \citenamefont {Schuster},
  \citenamefont {Frunzio}, \citenamefont {Devoret}, \citenamefont {Girvin},\
  and\ \citenamefont {Schoelkopf}}]{Houck2008}%
  \BibitemOpen
  \bibfield  {author} {\bibinfo {author} {\bibfnamefont {A.~A.}\ \bibnamefont
  {Houck}}, \bibinfo {author} {\bibfnamefont {J.~A.}\ \bibnamefont {Schreier}},
  \bibinfo {author} {\bibfnamefont {B.~R.}\ \bibnamefont {Johnson}}, \bibinfo
  {author} {\bibfnamefont {J.~M.}\ \bibnamefont {Chow}}, \bibinfo {author}
  {\bibfnamefont {J.}~\bibnamefont {Koch}}, \bibinfo {author} {\bibfnamefont
  {J.~M.}\ \bibnamefont {Gambetta}}, \bibinfo {author} {\bibfnamefont {D.~I.}\
  \bibnamefont {Schuster}}, \bibinfo {author} {\bibfnamefont {L.}~\bibnamefont
  {Frunzio}}, \bibinfo {author} {\bibfnamefont {M.~H.}\ \bibnamefont
  {Devoret}}, \bibinfo {author} {\bibfnamefont {S.~M.}\ \bibnamefont {Girvin}},
  \ and\ \bibinfo {author} {\bibfnamefont {R.~J.}\ \bibnamefont {Schoelkopf}},\
  }\href {\doibase 10.1103/PhysRevLett.101.080502} {\bibfield  {journal}
  {\bibinfo  {journal} {Phys. Rev. Lett.}\ }\textbf {\bibinfo {volume} {101}},\
  \bibinfo {pages} {080502} (\bibinfo {year} {2008})}\BibitemShut {NoStop}%
\bibitem [{\citenamefont {Chang}\ \emph {et~al.}(2013)\citenamefont {Chang},
  \citenamefont {Vissers}, \citenamefont {C{\'o}rcoles}, \citenamefont
  {Sandberg}, \citenamefont {Gao}, \citenamefont {Abraham}, \citenamefont
  {Chow}, \citenamefont {Gambetta}, \citenamefont {Beth~Rothwell},
  \citenamefont {Keefe}, \citenamefont {Steffen},\ and\ \citenamefont
  {Pappas}}]{Chang2013}%
  \BibitemOpen
  \bibfield  {author} {\bibinfo {author} {\bibfnamefont {J.~B.}\ \bibnamefont
  {Chang}}, \bibinfo {author} {\bibfnamefont {M.~R.}\ \bibnamefont {Vissers}},
  \bibinfo {author} {\bibfnamefont {A.~D.}\ \bibnamefont {C{\'o}rcoles}},
  \bibinfo {author} {\bibfnamefont {M.}~\bibnamefont {Sandberg}}, \bibinfo
  {author} {\bibfnamefont {J.}~\bibnamefont {Gao}}, \bibinfo {author}
  {\bibfnamefont {D.~W.}\ \bibnamefont {Abraham}}, \bibinfo {author}
  {\bibfnamefont {J.~M.}\ \bibnamefont {Chow}}, \bibinfo {author}
  {\bibfnamefont {J.~M.}\ \bibnamefont {Gambetta}}, \bibinfo {author}
  {\bibfnamefont {M.}~\bibnamefont {Beth~Rothwell}}, \bibinfo {author}
  {\bibfnamefont {G.~A.}\ \bibnamefont {Keefe}}, \bibinfo {author}
  {\bibfnamefont {M.}~\bibnamefont {Steffen}}, \ and\ \bibinfo {author}
  {\bibfnamefont {D.~P.}\ \bibnamefont {Pappas}},\ }\href {\doibase
  http://dx.doi.org/10.1063/1.4813269} {\bibfield  {journal} {\bibinfo
  {journal} {Applied Physics Letters}\ }\textbf {\bibinfo {volume} {103}},\
  \bibinfo {eid} {012602} (\bibinfo {year} {2013})}\BibitemShut {NoStop}%
\bibitem [{\citenamefont {Barends}\ \emph {et~al.}(2013)\citenamefont
  {Barends}, \citenamefont {Kelly}, \citenamefont {Megrant}, \citenamefont
  {Sank}, \citenamefont {Jeffrey}, \citenamefont {Chen}, \citenamefont {Yin},
  \citenamefont {Chiaro}, \citenamefont {Mutus}, \citenamefont {Neill},
  \citenamefont {O'Malley}, \citenamefont {Roushan}, \citenamefont {Wenner},
  \citenamefont {White}, \citenamefont {Cleland},\ and\ \citenamefont
  {Martinis}}]{Barends2013}%
  \BibitemOpen
  \bibfield  {author} {\bibinfo {author} {\bibfnamefont {R.}~\bibnamefont
  {Barends}}, \bibinfo {author} {\bibfnamefont {J.}~\bibnamefont {Kelly}},
  \bibinfo {author} {\bibfnamefont {A.}~\bibnamefont {Megrant}}, \bibinfo
  {author} {\bibfnamefont {D.}~\bibnamefont {Sank}}, \bibinfo {author}
  {\bibfnamefont {E.}~\bibnamefont {Jeffrey}}, \bibinfo {author} {\bibfnamefont
  {Y.}~\bibnamefont {Chen}}, \bibinfo {author} {\bibfnamefont {Y.}~\bibnamefont
  {Yin}}, \bibinfo {author} {\bibfnamefont {B.}~\bibnamefont {Chiaro}},
  \bibinfo {author} {\bibfnamefont {J.}~\bibnamefont {Mutus}}, \bibinfo
  {author} {\bibfnamefont {C.}~\bibnamefont {Neill}}, \bibinfo {author}
  {\bibfnamefont {P.}~\bibnamefont {O'Malley}}, \bibinfo {author}
  {\bibfnamefont {P.}~\bibnamefont {Roushan}}, \bibinfo {author} {\bibfnamefont
  {J.}~\bibnamefont {Wenner}}, \bibinfo {author} {\bibfnamefont {T.~C.}\
  \bibnamefont {White}}, \bibinfo {author} {\bibfnamefont {A.~N.}\ \bibnamefont
  {Cleland}}, \ and\ \bibinfo {author} {\bibfnamefont {J.~M.}\ \bibnamefont
  {Martinis}},\ }\href {\doibase 10.1103/PhysRevLett.111.080502} {\bibfield
  {journal} {\bibinfo  {journal} {Phys. Rev. Lett.}\ }\textbf {\bibinfo
  {volume} {111}},\ \bibinfo {pages} {080502} (\bibinfo {year}
  {2013})}\BibitemShut {NoStop}%
\bibitem [{\citenamefont {Wang}\ \emph {et~al.}(2009)\citenamefont {Wang},
  \citenamefont {Hofheinz}, \citenamefont {Wenner}, \citenamefont {Ansmann},
  \citenamefont {Bialczak}, \citenamefont {Lenander}, \citenamefont {Lucero},
  \citenamefont {Neeley}, \citenamefont {{O{\textquoteright}Connell}},
  \citenamefont {Sank}, \citenamefont {Weides}, \citenamefont {Cleland},\ and\
  \citenamefont {Martinis}}]{wang09a}%
  \BibitemOpen
  \bibfield  {author} {\bibinfo {author} {\bibfnamefont {H.}~\bibnamefont
  {Wang}}, \bibinfo {author} {\bibfnamefont {M.}~\bibnamefont {Hofheinz}},
  \bibinfo {author} {\bibfnamefont {J.}~\bibnamefont {Wenner}}, \bibinfo
  {author} {\bibfnamefont {M.}~\bibnamefont {Ansmann}}, \bibinfo {author}
  {\bibfnamefont {R.~C.}\ \bibnamefont {Bialczak}}, \bibinfo {author}
  {\bibfnamefont {M.}~\bibnamefont {Lenander}}, \bibinfo {author}
  {\bibfnamefont {E.}~\bibnamefont {Lucero}}, \bibinfo {author} {\bibfnamefont
  {M.}~\bibnamefont {Neeley}}, \bibinfo {author} {\bibfnamefont {A.~D.}\
  \bibnamefont {{O{\textquoteright}Connell}}}, \bibinfo {author} {\bibfnamefont
  {D.}~\bibnamefont {Sank}}, \bibinfo {author} {\bibfnamefont {M.}~\bibnamefont
  {Weides}}, \bibinfo {author} {\bibfnamefont {A.~N.}\ \bibnamefont {Cleland}},
  \ and\ \bibinfo {author} {\bibfnamefont {J.~M.}\ \bibnamefont {Martinis}},\
  }\href {\doibase doi:10.1063/1.3273372} {\bibfield  {journal} {\bibinfo
  {journal} {Applied Physics Letters}\ }\textbf {\bibinfo {volume} {95}},\
  \bibinfo {pages} {233508} (\bibinfo {year} {2009})}\BibitemShut {NoStop}%
\bibitem [{\citenamefont {Geerlings}\ \emph {et~al.}(2012)\citenamefont
  {Geerlings}, \citenamefont {Shankar}, \citenamefont {Edwards}, \citenamefont
  {Frunzio}, \citenamefont {Schoelkopf},\ and\ \citenamefont
  {Devoret}}]{Geerlings2012}%
  \BibitemOpen
  \bibfield  {author} {\bibinfo {author} {\bibfnamefont {K.}~\bibnamefont
  {Geerlings}}, \bibinfo {author} {\bibfnamefont {S.}~\bibnamefont {Shankar}},
  \bibinfo {author} {\bibfnamefont {E.}~\bibnamefont {Edwards}}, \bibinfo
  {author} {\bibfnamefont {L.}~\bibnamefont {Frunzio}}, \bibinfo {author}
  {\bibfnamefont {R.~J.}\ \bibnamefont {Schoelkopf}}, \ and\ \bibinfo {author}
  {\bibfnamefont {M.~H.}\ \bibnamefont {Devoret}},\ }\href {\doibase
  http://dx.doi.org/10.1063/1.4710520} {\bibfield  {journal} {\bibinfo
  {journal} {Applied Physics Letters}\ }\textbf {\bibinfo {volume} {100}},\
  \bibinfo {eid} {192601} (\bibinfo {year} {2012})}\BibitemShut {NoStop}%
\bibitem [{\citenamefont {Gao}\ \emph {et~al.}(2008)\citenamefont {Gao},
  \citenamefont {Daal}, \citenamefont {Vayonakis}, \citenamefont {Kumar},
  \citenamefont {Zmuidzinas}, \citenamefont {Sadoulet}, \citenamefont {Mazin},
  \citenamefont {Day},\ and\ \citenamefont {Leduc}}]{gao08}%
  \BibitemOpen
  \bibfield  {author} {\bibinfo {author} {\bibfnamefont {J.}~\bibnamefont
  {Gao}}, \bibinfo {author} {\bibfnamefont {M.}~\bibnamefont {Daal}}, \bibinfo
  {author} {\bibfnamefont {A.}~\bibnamefont {Vayonakis}}, \bibinfo {author}
  {\bibfnamefont {S.}~\bibnamefont {Kumar}}, \bibinfo {author} {\bibfnamefont
  {J.}~\bibnamefont {Zmuidzinas}}, \bibinfo {author} {\bibfnamefont
  {B.}~\bibnamefont {Sadoulet}}, \bibinfo {author} {\bibfnamefont {B.~A.}\
  \bibnamefont {Mazin}}, \bibinfo {author} {\bibfnamefont {P.~K.}\ \bibnamefont
  {Day}}, \ and\ \bibinfo {author} {\bibfnamefont {H.~G.}\ \bibnamefont
  {Leduc}},\ }\href {\doibase doi:10.1063/1.2906373} {\bibfield  {journal}
  {\bibinfo  {journal} {Applied Physics Letters}\ }\textbf {\bibinfo {volume}
  {92}},\ \bibinfo {pages} {152505} (\bibinfo {year} {2008})}\BibitemShut
  {NoStop}%
\bibitem [{\citenamefont {Wenner}\ \emph {et~al.}(2011)\citenamefont {Wenner},
  \citenamefont {Neeley}, \citenamefont {Bialczak}, \citenamefont {Lenander},
  \citenamefont {Lucero}, \citenamefont {{O{\textquoteright}Connell}},
  \citenamefont {Sank}, \citenamefont {Wang}, \citenamefont {Weides},
  \citenamefont {Cleland},\ and\ \citenamefont {Martinis}}]{wenner11}%
  \BibitemOpen
  \bibfield  {author} {\bibinfo {author} {\bibfnamefont {J.}~\bibnamefont
  {Wenner}}, \bibinfo {author} {\bibfnamefont {M.}~\bibnamefont {Neeley}},
  \bibinfo {author} {\bibfnamefont {R.~C.}\ \bibnamefont {Bialczak}}, \bibinfo
  {author} {\bibfnamefont {M.}~\bibnamefont {Lenander}}, \bibinfo {author}
  {\bibfnamefont {E.}~\bibnamefont {Lucero}}, \bibinfo {author} {\bibfnamefont
  {A.~D.}\ \bibnamefont {{O{\textquoteright}Connell}}}, \bibinfo {author}
  {\bibfnamefont {D.}~\bibnamefont {Sank}}, \bibinfo {author} {\bibfnamefont
  {H.}~\bibnamefont {Wang}}, \bibinfo {author} {\bibfnamefont {M.}~\bibnamefont
  {Weides}}, \bibinfo {author} {\bibfnamefont {A.~N.}\ \bibnamefont {Cleland}},
  \ and\ \bibinfo {author} {\bibfnamefont {J.~M.}\ \bibnamefont {Martinis}},\
  }\href {\doibase 10.1088/0953-2048/24/6/065001} {\bibfield  {journal}
  {\bibinfo  {journal} {Superconductor Science and Technology}\ }\textbf
  {\bibinfo {volume} {24}},\ \bibinfo {pages} {065001} (\bibinfo {year}
  {2011})}\BibitemShut {NoStop}%
\bibitem [{\citenamefont {Quintana}\ \emph {et~al.}(2014)\citenamefont
  {Quintana}, \citenamefont {Megrant}, \citenamefont {Chen}, \citenamefont
  {Dunsworth}, \citenamefont {Chiaro}, \citenamefont {Barends}, \citenamefont
  {Campbell}, \citenamefont {Chen}, \citenamefont {Hoi}, \citenamefont
  {Jeffrey}, \citenamefont {Kelly}, \citenamefont {Mutus}, \citenamefont
  {{O'Malley}}, \citenamefont {Neill}, \citenamefont {Roushan}, \citenamefont
  {Sank}, \citenamefont {Vainsencher}, \citenamefont {Wenner}, \citenamefont
  {White}, \citenamefont {Cleland},\ and\ \citenamefont
  {Martinis}}]{quintana14}%
  \BibitemOpen
  \bibfield  {author} {\bibinfo {author} {\bibfnamefont {C.~M.}\ \bibnamefont
  {Quintana}}, \bibinfo {author} {\bibfnamefont {A.}~\bibnamefont {Megrant}},
  \bibinfo {author} {\bibfnamefont {Z.}~\bibnamefont {Chen}}, \bibinfo {author}
  {\bibfnamefont {A.}~\bibnamefont {Dunsworth}}, \bibinfo {author}
  {\bibfnamefont {B.}~\bibnamefont {Chiaro}}, \bibinfo {author} {\bibfnamefont
  {R.}~\bibnamefont {Barends}}, \bibinfo {author} {\bibfnamefont
  {B.}~\bibnamefont {Campbell}}, \bibinfo {author} {\bibfnamefont
  {Y.}~\bibnamefont {Chen}}, \bibinfo {author} {\bibfnamefont {I.}~\bibnamefont
  {Hoi}}, \bibinfo {author} {\bibfnamefont {E.}~\bibnamefont {Jeffrey}},
  \bibinfo {author} {\bibfnamefont {J.}~\bibnamefont {Kelly}}, \bibinfo
  {author} {\bibfnamefont {J.~Y.}\ \bibnamefont {Mutus}}, \bibinfo {author}
  {\bibfnamefont {P.~J.~J.}\ \bibnamefont {{O'Malley}}}, \bibinfo {author}
  {\bibfnamefont {C.}~\bibnamefont {Neill}}, \bibinfo {author} {\bibfnamefont
  {P.}~\bibnamefont {Roushan}}, \bibinfo {author} {\bibfnamefont
  {D.}~\bibnamefont {Sank}}, \bibinfo {author} {\bibfnamefont {A.}~\bibnamefont
  {Vainsencher}}, \bibinfo {author} {\bibfnamefont {J.}~\bibnamefont {Wenner}},
  \bibinfo {author} {\bibfnamefont {T.~C.}\ \bibnamefont {White}}, \bibinfo
  {author} {\bibfnamefont {A.~N.}\ \bibnamefont {Cleland}}, \ and\ \bibinfo
  {author} {\bibfnamefont {J.~M.}\ \bibnamefont {Martinis}},\ }\href
  {http://arxiv.org/abs/1407.4769} {\bibfield  {journal} {\bibinfo  {journal}
  {{arXiv:1407.4769} [cond-mat, physics:quant-ph]}\ } (\bibinfo {year}
  {2014})}\BibitemShut {NoStop}%
\bibitem [{\citenamefont {Paik}\ \emph {et~al.}(2011)\citenamefont {Paik},
  \citenamefont {Schuster}, \citenamefont {Bishop}, \citenamefont {Kirchmair},
  \citenamefont {Catelani}, \citenamefont {Sears}, \citenamefont {Johnson},
  \citenamefont {Reagor}, \citenamefont {Frunzio}, \citenamefont {Glazman},
  \citenamefont {Girvin}, \citenamefont {Devoret},\ and\ \citenamefont
  {Schoelkopf}}]{paik11}%
  \BibitemOpen
  \bibfield  {author} {\bibinfo {author} {\bibfnamefont {H.}~\bibnamefont
  {Paik}}, \bibinfo {author} {\bibfnamefont {D.~I.}\ \bibnamefont {Schuster}},
  \bibinfo {author} {\bibfnamefont {L.~S.}\ \bibnamefont {Bishop}}, \bibinfo
  {author} {\bibfnamefont {G.}~\bibnamefont {Kirchmair}}, \bibinfo {author}
  {\bibfnamefont {G.}~\bibnamefont {Catelani}}, \bibinfo {author}
  {\bibfnamefont {A.~P.}\ \bibnamefont {Sears}}, \bibinfo {author}
  {\bibfnamefont {B.~R.}\ \bibnamefont {Johnson}}, \bibinfo {author}
  {\bibfnamefont {M.~J.}\ \bibnamefont {Reagor}}, \bibinfo {author}
  {\bibfnamefont {L.}~\bibnamefont {Frunzio}}, \bibinfo {author} {\bibfnamefont
  {L.~I.}\ \bibnamefont {Glazman}}, \bibinfo {author} {\bibfnamefont {S.~M.}\
  \bibnamefont {Girvin}}, \bibinfo {author} {\bibfnamefont {M.~H.}\
  \bibnamefont {Devoret}}, \ and\ \bibinfo {author} {\bibfnamefont {R.~J.}\
  \bibnamefont {Schoelkopf}},\ }\href {\doibase 10.1103/PhysRevLett.107.240501}
  {\bibfield  {journal} {\bibinfo  {journal} {Physical Review Letters}\
  }\textbf {\bibinfo {volume} {107}},\ \bibinfo {pages} {240501} (\bibinfo
  {year} {2011})}\BibitemShut {NoStop}%
\bibitem [{\citenamefont {Rigetti}\ \emph {et~al.}(2012)\citenamefont
  {Rigetti}, \citenamefont {Gambetta}, \citenamefont {Poletto}, \citenamefont
  {Plourde}, \citenamefont {Chow}, \citenamefont {C\'{o}rcoles}, \citenamefont
  {Smolin}, \citenamefont {Merkel}, \citenamefont {Rozen}, \citenamefont
  {Keefe}, \citenamefont {Rothwell}, \citenamefont {Ketchen},\ and\
  \citenamefont {Steffen}}]{rigetti12}%
  \BibitemOpen
  \bibfield  {author} {\bibinfo {author} {\bibfnamefont {C.}~\bibnamefont
  {Rigetti}}, \bibinfo {author} {\bibfnamefont {J.~M.}\ \bibnamefont
  {Gambetta}}, \bibinfo {author} {\bibfnamefont {S.}~\bibnamefont {Poletto}},
  \bibinfo {author} {\bibfnamefont {B.~L.~T.}\ \bibnamefont {Plourde}},
  \bibinfo {author} {\bibfnamefont {J.~M.}\ \bibnamefont {Chow}}, \bibinfo
  {author} {\bibfnamefont {A.~D.}\ \bibnamefont {C\'{o}rcoles}}, \bibinfo
  {author} {\bibfnamefont {J.~A.}\ \bibnamefont {Smolin}}, \bibinfo {author}
  {\bibfnamefont {S.~T.}\ \bibnamefont {Merkel}}, \bibinfo {author}
  {\bibfnamefont {J.~R.}\ \bibnamefont {Rozen}}, \bibinfo {author}
  {\bibfnamefont {G.~A.}\ \bibnamefont {Keefe}}, \bibinfo {author}
  {\bibfnamefont {M.~B.}\ \bibnamefont {Rothwell}}, \bibinfo {author}
  {\bibfnamefont {M.~B.}\ \bibnamefont {Ketchen}}, \ and\ \bibinfo {author}
  {\bibfnamefont {M.}~\bibnamefont {Steffen}},\ }\href {\doibase
  10.1103/PhysRevB.86.100506} {\bibfield  {journal} {\bibinfo  {journal}
  {Physical Review B}\ }\textbf {\bibinfo {volume} {86}},\ \bibinfo {pages}
  {100506} (\bibinfo {year} {2012})}\BibitemShut {NoStop}%
\bibitem [{\citenamefont {Creedon}\ \emph {et~al.}(2011)\citenamefont
  {Creedon}, \citenamefont {Reshitnyk}, \citenamefont {Farr}, \citenamefont
  {Martinis}, \citenamefont {Duty},\ and\ \citenamefont {Tobar}}]{creedon11}%
  \BibitemOpen
  \bibfield  {author} {\bibinfo {author} {\bibfnamefont {D.~L.}\ \bibnamefont
  {Creedon}}, \bibinfo {author} {\bibfnamefont {Y.}~\bibnamefont {Reshitnyk}},
  \bibinfo {author} {\bibfnamefont {W.}~\bibnamefont {Farr}}, \bibinfo {author}
  {\bibfnamefont {J.~M.}\ \bibnamefont {Martinis}}, \bibinfo {author}
  {\bibfnamefont {T.~L.}\ \bibnamefont {Duty}}, \ and\ \bibinfo {author}
  {\bibfnamefont {M.~E.}\ \bibnamefont {Tobar}},\ }\href {\doibase
  doi:10.1063/1.3595942} {\bibfield  {journal} {\bibinfo  {journal} {Applied
  Physics Letters}\ }\textbf {\bibinfo {volume} {98}},\ \bibinfo {pages}
  {222903} (\bibinfo {year} {2011})}\BibitemShut {NoStop}%
\bibitem [{\citenamefont {Chavez}\ and\ \citenamefont {Hess}(2003)}]{chavez03}%
  \BibitemOpen
  \bibfield  {author} {\bibinfo {author} {\bibfnamefont {K.~L.}\ \bibnamefont
  {Chavez}}\ and\ \bibinfo {author} {\bibfnamefont {D.~W.}\ \bibnamefont
  {Hess}},\ }\href {\doibase 10.1149/1.1557085} {\bibfield  {journal} {\bibinfo
   {journal} {Journal of The Electrochemical Society}\ }\textbf {\bibinfo
  {volume} {150}},\ \bibinfo {pages} {G284} (\bibinfo {year}
  {2003})}\BibitemShut {NoStop}%
\bibitem [{\citenamefont {Her}, \citenamefont {Beams},\ and\ \citenamefont
  {Novotny}(2013)}]{her13}%
  \BibitemOpen
  \bibfield  {author} {\bibinfo {author} {\bibfnamefont {M.}~\bibnamefont
  {Her}}, \bibinfo {author} {\bibfnamefont {R.}~\bibnamefont {Beams}}, \ and\
  \bibinfo {author} {\bibfnamefont {L.}~\bibnamefont {Novotny}},\ }\href
  {\doibase 10.1016/j.physleta.2013.04.015} {\bibfield  {journal} {\bibinfo
  {journal} {Physics Letters A}\ }\textbf {\bibinfo {volume} {377}},\ \bibinfo
  {pages} {1455} (\bibinfo {year} {2013})}\BibitemShut {NoStop}%
\end{thebibliography}%

\end{document}